\documentclass{article}
\usepackage{amsmath, amssymb, comment}
\newtheorem{theorem}{Theorem}[section]
\newtheorem{corollary}[theorem]{Corollary}
\newtheorem{lemma}[theorem]{Lemma}

\newtheorem{definition}[theorem]{Definition}

\usepackage{wrapfig}
\usepackage{tikz}
\usetikzlibrary{arrows,snakes,backgrounds}
\usetikzlibrary{calc}
\usepackage{relsize}

\usepackage{times}
\usepackage{graphicx} % more modern
\usepackage{subfigure} 

\usepackage{natbib}

\usepackage{algorithm}
\usepackage{algorithmic}

\usepackage{hyperref}

\usepackage[accepted]{icml2017}

\DeclareMathOperator{\tr}{tr}
\DeclareMathOperator{\E}{\mathbb{E}}

\icmltitlerunning{When can Multi-Site Datasets be Pooled for Regression?}

\begin{document} 
\twocolumn[
\icmltitle{When can Multi-Site Datasets be Pooled for Regression?\\
  Hypothesis Tests, $\ell_2$-consistency and Neuroscience Applications}



% you can specify symbols, otherwise they are numbered in order

\begin{icmlauthorlist}
\icmlauthor{Hao Henry Zhou}{UW}
\icmlauthor{Yilin Zhang}{UW}
\icmlauthor{Vamsi K. Ithapu}{UW}
\icmlauthor{Sterling C. Johnson}{UW,VA}
\icmlauthor{Grace Wahba}{UW}
\icmlauthor{Vikas Singh}{UW}
\end{icmlauthorlist}

\icmlaffiliation{UW}{University of Wisconsin-Madison}
\icmlaffiliation{VA}{William S. Middleton Memorial Veteran's Affairs Hospital}

\icmlcorrespondingauthor{Hao Zhou}{hzhou@stat.wisc.edu}
\icmlcorrespondingauthor{Vikas Singh}{vsingh@biostat.wisc.edu}

\icmlkeywords{Transfer learning, Multitask learning, Domain adaptation, Dataset shift, Multi-sites, Data harmonization}

\vskip 0.3in
]




\printAffiliationsAndNotice{} % otherwise use the standard text.

%\footnotetext{hi}

\begin{abstract} 
Many studies in biomedical and health sciences involve small sample sizes due to logistic or financial constraints. 
Often, identifying weak (but scientifically interesting) associations between a set of predictors and a response necessitates pooling datasets from multiple diverse labs or groups. 
While there is a rich literature in statistical machine learning to address distributional 
shifts and inference in multi-site datasets, it is less clear {\it when} such pooling is guaranteed to help (and when it does not) -- independent of the inference algorithms we use. In this paper, we present a hypothesis test to answer this question, both for classical and high dimensional linear regression. 
We precisely identify regimes where pooling datasets across multiple sites is sensible, and how such policy decisions can be made via simple checks executable on each site before any data transfer ever happens. 
With a focus on Alzheimer's disease studies, we present empirical results showing that in regimes suggested by our analysis, pooling a local dataset with data from an international study improves power. 
\end{abstract}

\section{Introduction}\label{sec:intro}
In the last two decades, statistical machine learning algorithms for processing massive datasets have been intensively studied for a wide-range of applications in computer vision, biology, chemistry and healthcare \cite{murdoch2013inevitable, tarca2007machine}.
While the challenges posed by large scale datasets are compelling, one is often faced with a fairly distinct set of technical issues for studies in biological and health sciences. For instance, a sizable portion of scientific research is carried out by small or medium-sized groups \cite{fortin13} supported by modest budgets \cite{mike16}. 
Hence, there are logistic/financial constraints on the number of experiments and/or number of participants within a trial, leading to small size datasets. While the analysis may be sufficiently powered to evaluate the {\em primary} hypothesis of the study/experiment,
interesting follow-up scientific questions (often more {\em nuanced}), come up during the course of the project. 
These tasks may be underpowered for the sample sizes available.
This necessitates efforts to identify similar datasets elsewhere so that the combined sample size of the ``pooled'' dataset is enough to determine significant associations between a response and a set of predictors, e.g., within linear regression. 

{\em Motivating Application.} In genomics, funding agencies %and consortia 
have invested effort into standardizing/curating data collection across large international projects \cite{encode2004encode}. 
In other disciplines, such as in the study of neurological disorders, heterogeneity in disease etiology, variations in scanners and/or acquisition tools make standardization more difficult. 
For Alzheimer's disease (AD), a motivation of this work, efforts such as ADNI \cite{weiner15} provide a variety of clinical, imaging and cognitive tests data for $800+$ {\em older adults}. 
However, the research focus has now moved to the early stages of disease -- as early as {\em late middle age} -- where treatments are expected to be more effective. 
But (a) the statistical signal at this stage is {\em weak} and difficult to demonstrate without large sample sizes and (b) such ``preclinical'' participants are not well represented, even in large ADNI sized studies. 
Hence, there is a concerted effort in general for smaller standalone projects (focused on a specific disease stage),
that can be retrospectively pooled for analysis towards addressing a challenging scientific hypothesis \cite{jahanshad13}. 
Unfortunately, acquisition protocols for various measures across sites are usually different and data are heterogeneous. 
These issues raise a fundamental technical question. 
When is it meaningful to pool datasets for estimating a simple statistical model
(e.g., linear regression)? When can we guarantee improvements in statistical power, and when are such pooling efforts not worth it?
Can we give a hypothesis test and obtain $p$-values to inform our policies/decisions? 
While related problems have been studied in machine learning from an algorithm design perspective, even simple hypothesis tests which can be deployed by a researcher in practice, are currently unavailable.
Our goal is to remove this significant limitation. 

{\em Putting our development in context.} 
The realization that ``similar'' datasets from multiple sites can be pooled to potentially improve statistical power is not new.
With varying empirical success, models tailored to perform regression in multi-site studies \cite{group02}, \cite{haase09}, \cite{klunk15} 
have been proposed, where due to operational reasons, recruitment and data acquisition are distributed over multiple sites, 
or even countries.
When the pooling is being performed {\em retrospectively} (i.e., after the data has been collected), resolving site-specific confounds, such as distributional shifts or biases in measurements, is essential before estimation/inference of a statistical model. 
We will {\em not} develop new algorithms for estimating the distributional mismatch or for performing multi-site regression --- 
rather, our primary goal is to identify the regimes (and give easily computable checks) where this regression task on a pooled dataset is statistically meaningful, assuming that good pre-processing schemes are available. We will present a rigorous yet simple to implement hypothesis test, analyze its behavior, and show extensive experimental evidence (for an important scientific problem). 
The practitioner is free to use his/her preferred procedure for the ``before step'' (estimating the distributional shifts). 

{\bf Contributions:}
{\bf a)} Our main result is a hypothesis test to evaluate whether pooling
data across multiple sites for regression (before or after correcting for site-specific distributional shifts)
can improve the estimation (mean squared error) of the relevant coefficients (while permitting
an influence from a set of confounding variables).
{\bf b)} We derive analogous results in the high-dimensional setting by leveraging
a different set of analysis techniques. Using an existing sparse multi-task Lasso model, we show how the utility
of pooling can be evaluated even when the support set of the features (predictors) is not exactly the same across sites
using ideas broadly related to high dimensional simultaneous inference \cite{dezeure15}. We show $\ell_2$-consistency rate, which supports the use of sparse multi-task Lasso when sparsity patterns are not totally identical. 
{\bf c)} On an important scientific problem of analyzing early Alzheimer's disease (AD) individuals, we provide compelling experimental results showing consistent acceptance rate and statistical power.
Via a publicly available software package, this will facilitate many multi-site regression analysis efforts in the short to medium
term future. 

\subsection{Related Work}
{\bf Meta-analysis approaches.} 
If datasets at multiple different sites {\em cannot} be shared or pooled, the task of deriving meaningful scientific 
conclusions from {\em results of multiple independently conducted analyses} generally falls under the umbrella 
term of ``meta analysis''. 
The literature provides various strategies to cumulate
the general findings from analyses on different datasets. But even experts believe that, minor violations of assumptions can lead to misleading scientific conclusions \cite{greco13}, and substantial personal judgment (and expertise) is needed to conduct them. 
It is widely accepted that when the ability to pool the data is an option, simpler schemes may perform better. 

{\bf Domain adaptation/shift.} Separately, the idea of addressing ``shift'' within datasets has been rigorously studied within statistical 
machine learning, see \cite{patel15,li12}. 
For example, domain adaptation, including dataset and covariate shift, seeks to align (the distributions of) multiple datasets to enable follow-up processing \cite{ben2003exploiting}. 
Typically, such algorithms assume a bias in the sampling process, and adopt re-weighting as the solution \cite{huang07,gong13}. 
Alternatively, a family of such methods assume that sites (or datasets) differ due to feature distortions (e.g., calibration error),
which are resolved, in general, by minimizing some distance measure between appropriate distributions \cite{bakt13,pan11,long15,ganin16}.
In general, these approaches have nice theoretical properties \cite{ben10,cortes11,zhou16}.
However, it is important to note that the domain adaptation literature {\em focuses on the algorithm itself} -- to resolve the distributional site-wise differences. 
It does {\em not} address the issue of whether pooling the datasets, after applying the calculated adaptation (i.e., transformation), is beneficial. 
Our goal in this work is to assess whether multiple datasets can be pooled --- either {\em before} or usually {\em after}
applying the best domain adaptation methods --- for improving our estimation of the relevant coefficients within linear regression.
We propose a hypothesis test to directly address this question. 

{\bf The high-dimensional case.} 
High dimensional scenarios, in general, involve predicting a response (e.g., cognitive score) from high dimensional predictors such as image scans (or derived features) and genetic data, which in general, entails Lasso-type formulations unlike the classical regression models.
Putting multi-task representation learning \cite{maurer16,ando05,maurer13} together with a sparsity regularizer, we get the multi-task Lasso model \cite{liuml09,kim10}.
Although this seems like a suitable model \cite{chen12}, it assumes that the multiple tasks (sites here) have an identical active set of predictors.
Instead, we find that the sparse multi-task Lasso \cite{lee10}, roughly, a multi-task version of sparse group Lasso \cite{simon13,lee10} is a better starting point.
There is no theoretical analysis in \cite{simon13}; although a $\ell_2$-consistency for sparse group lasso is derived in \cite{chatterjee12} using a general proof procedure for M-estimators, it does not take into account the specific sparse group Lasso properties. This makes the result non-informative for sparse group Lasso (much less, sparse multi-task Lasso).
Specifically, as we will see shortly, in sparse multi-task Lasso, the joint effects of two penalties induces a special type of asymmetric structure. 
We show a new result, in the style of Lasso \cite{mein09,liugl09}, for $\ell_2$ convergence rate for this model.
It matches known results for Lasso/group Lasso, and identifies regimes where the sparse multi-task (multi-site) setting is advantageous.

{\bf Simultaneous High dimensional Inference.}
Simultaneous high dimensional inference models such as multi sample-splitting and de-biased Lasso is an active research topic in statistics \cite{dezeure15}. 
Multi sample-splitting use half of the dataset for variable selection and the rest for calculating $p$-values. 
De-biased Lasso chooses one feature as a response and the others as predictors to estimate a Lasso model; this procedure is repeated for each feature. 
Estimators from De-biased Lasso asymptotically follow the multi-normal distribution \cite{dezeure2016high}, and using Bonferroni-Holm adjustment produces simultaneous $p$-values. 
Such ideas together with the $\ell_2$-convergence results for sparse multitask Lasso, will help extend our analysis to the high dimensional setting.

\section{Hypothesis Test for Multi-Site Regression} \label{sec:hyptest}

We first describe a simple setting where one seeks to apply standard linear regression to data pooled from multiple sites. 
For presentation purposes, we will deal with variable selection issues later. Within this setup, we will introduce our main result --- a hypothesis test to evaluate statistical power improvements (e.g., mean squared error) when running a regression model on a pooled dataset. 
We will see that the proposed test is transparent to the use of adaptation algorithms, if any, to pre-process the multi-site data (more details in appendix).
In later sections, we will present convergence analysis and extensions to the large $p$ setting.
Matrices (vectors/scalars) are upper case (and lower case). $\lVert .\rVert_{\ast}$ is the nuclear norm.  

We first introduce the single-site regression model.
Let $X \in \mathbb{R}^{n\times p}$ and $y \in \mathbb{R}^{n\times 1}$ denote the feature matrix of predictors and the response vector respectively.
If $\beta$ corresponds to the coefficient vector (i.e., predictor weights), then the regression model is 
\begin{align}\label{Onesetmodel}
  \min_{\beta}\quad & \frac{1}{n}\lVert y-X\beta\rVert^2_2
\end{align}
where $y = X\beta^{\ast} + \epsilon$ and $\epsilon\sim N(0,\sigma^2)\ \textsc{i.i.d.}$ if 
$\beta^{\ast}$ is the true coefficient vector from which $y$ is generated.
The mean-squared error (MSE) and $\ell_2$-consistency of regression is well-known. The mean-squared error (MSE) of \eqref{Onesetmodel} is
$\E\lVert\hat{\beta}-\beta^{\ast}\rVert^2_2 = \tr\left((X^TX)^{-1}\right)\sigma^2$. If $k$ denotes the number of sites, then one may first apply a domain adaptation scheme to
account for the distributional shifts between the $k$ different predictors $\{X_i\}_{i=1}^k$, and then run a regression model.
If the underlying ``concept'' (i.e., predictors and responses relationship)
can be assumed to be the same across the different sites, then it is reasonable to impose the {\it same} $\beta$ for all sites. 
For instance, the influence of CSF protein measurements on cognitive scores of an individual may be invariant to demographics. 
Nonetheless, if the distributional mismatch correction is imperfect, 
we may define $\Delta \beta_i = \beta_i - \beta^{\ast}$ where $i \in \{1,\ldots,k\}$ as the residual difference between the site-specific coefficients 
and the true shared coefficient vector (in the ideal case, we have $\Delta \beta_i=0$).
In the multi-site setting, we can write 
\begin{align}\label{Ksetsmodel-total}
  \min_{\beta} \sum_{i=1}^k \tau^2_i\lVert y_i-X_i\beta\rVert^2_2
\end{align}
where for each site $i$ we have $y_i = X_i\beta^{\ast} + X_i\Delta \beta_i +\epsilon_i$ 
and $\epsilon_ i\sim \mathcal{N}(0,\sigma^2_i) \quad \textsc{i.i.d}$. Here, $\tau_i$ is a weighting parameter for each site, 
if such information is available.

Our main goal is to test if the combined regression improves the estimation for a single site.
We can pose this question in terms of improvements in the mean squared error (MSE). 
%{\color{blue} Hence, \textsc{w.l.o.g.} using site $1$ as the reference, we have the following reduced objective by setting $\beta_1=\beta^{\ast}$ and $\tau_1=1$ in \eqref{Ksetsmodel-total}, }
Hence, \textsc{w.l.o.g.} using site $1$ as the reference, we set $\tau_1=1$ in \eqref{Ksetsmodel-total} and consider $\beta^{\ast}=\beta_1$,
\begin{align}\label{Ksetsmodel}
\min_{\beta} \lVert y_1-X_1\beta\rVert^2_2 + \sum_{i=2}^k \tau^2_i\lVert y_i-X_i\beta\rVert^2_2
  \end{align}
\begin{wrapfigure}[14]{rb}{0pt}
%	\begin{figure}[H]
%		\vskip -0.05in
%		\begin{center}
			\begin{tikzpicture}
				\filldraw[black!40!green]
			(0,0) circle (1pt) node[align=center,  above](betahat){};   
			\filldraw[red]
			(-0.4,0) circle (1pt) node[align=center,  below, red,
			font=\fontsize{10}{10}\selectfont] { $\beta_1$};
			\filldraw[blue]
			(0.4,0) circle (1pt) node[align=right,  below, blue,
			font=\fontsize{10}{10}\selectfont] {$\beta_2$};
			
			\node[black!40!green] at (betahat.north west) {$\hat{\beta}$};
			\draw[very thick, dotted,fill=black!40!green, opacity=0.2] (0,0) circle (0.7cm);
			\draw[very thick, red] (-0.4,0) circle (1cm);
			\draw[very thick, blue] (0.4,0) circle (1cm);
			\end{tikzpicture}
			%\vspace{-1cm}
		        \caption{\footnotesize \label{Explanationtradeoff} $\beta_1$ and $\beta_2$ are $1^{st}$ and $2^{nd}$ site coefficients. 
                          After combination, $\beta_1$'s bias increases but variance reduces, resulting in a smaller MSE.}
%			\end{center}
%		\vskip -0.2in
%	\end{figure}
\end{wrapfigure}
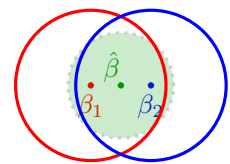

Clearly, when the sample size is not large enough, the multi-site formulation in \eqref{Ksetsmodel} may reduce variance significantly, because of the averaging effect in the objective function, while increasing the bias by a little. This reduces the Mean Squared Error (MSE), see Figure \ref{Explanationtradeoff}. 
Note that while traditionally, the unbiasedness property was desirable, an extensive body of literature on ridge regression suggests that the quantity of interest should really be $\E\|\hat{\beta} - \beta^\ast\|^2_2$ \cite{hoerl70,james61}.
These ideas are nicely studied within papers devoted to the ``bias-variance'' trade-off. 
Similar to these results, we will focus on the mean squared error because the asymptotic consistency properties that come with an unbiased estimator are not meaningful here anyway --- the key reason we want to pool datasets in the first place is because of small sample sizes. 
We now provide a result showing how the tuning parameters $\tau_2,\ldots,\tau_k$ can be chosen. 
\begin{theorem}\label{theorem:optimaltau} $\tau_i=\frac{\sigma_1}{\sigma_i}$ achieves the smallest variance in $\hat{\beta}$. \end{theorem}
\noindent {\it Remarks:} This result follows from observing that the each site's contribution is inversely proportional to site-specific noise level, $\sigma_i$.
We will show that this choice of $\tau_i$s also leads to a simple mechanism to setup a hypothesis test.

%%%%%%%%%%%%%%%%%%%%%%%%%%%%%%%%%%%%%%%%%%%%%%%%%%%%%%%%%%%%%%%%%%%%%%%%%%%%%%%%%%%%%%%%%%%%%%%%%%%%%%%%%%%%%%%%%%%%%%%%

\subsection{Sharing all \mathversion{bold}$\beta$s\mathversion{bold}}\label{sec:all}

In the specification above, the estimates of $\beta_i$ across all $k$ sites are restricted to be the same.
Without this constraint, \eqref{Ksetsmodel} is equivalent to fitting a regression {\it separately} on each site. 
So, a natural question is whether this constraint improves estimation. 
To evaluate whether MSE is reduced, we first need to quantify the change in the bias and variance of \eqref{Ksetsmodel} compared to \eqref{Onesetmodel}. 
To do so, we introduce a few notations. Let $n_i$ be the sample size of site $i$, and 
let $\hat{\beta}_i$ denote the regression estimate from a specific site $i$. We have 
$\Delta\hat{\beta}_i = \hat{\beta}_i-\hat{\beta}_1$.
We define the length $kp$ vector $\Delta \beta^T$ as $\Delta \beta^T = (\Delta\beta^T_2,...,\Delta\beta^T_k)$ (similarly for $\Delta \hat{\beta}^T$). 
We use $\hat{\Sigma}_i$ for the sample covariance matrix of the data (predictors) from the site $i$ and 
$G \in \mathbb{R}^{(k-1)p\times (k-1)p}$ for the covariance matrix of $\Delta \hat{\beta}$, 
where $G_{ii} = (n_1\hat{\Sigma}_1)^{-1} + (n_i\tau^2_i\hat{\Sigma}_i)^{-1}$ and $G_{ij}=(n_1\hat{\Sigma}_1)^{-1}$ whenever $i\neq j$. 

Let the difference in bias and variance between the single site model in \eqref{Onesetmodel} and the multi-site model in \eqref{Ksetsmodel} be $Bias_{\beta}$ and $Var_{\beta}$ respectively.
%denote the difference in variance.
Let $\hat{\Sigma}_2^k = \sum_{i=2}^k n_i\tau^2_i \hat{\Sigma}_i$ and $\hat{\Sigma}_1^k=n_1\hat{\Sigma}_1+\hat{\Sigma}_2^k$. We have, 
\begin{lemma}
For model \eqref{Ksetsmodel}, we have
\begin{small}\begin{align}\label{biasupper}
&\frac{\lVert Bias_{\beta}\rVert^2_2}{\lVert G^{-1/2}\Delta\beta\rVert^2_2} \leq\lVert(\hat{\Sigma}_1^k)^{-2}(\hat{\Sigma}_2^k (n_1\hat{\Sigma}_1)^{-1} \hat{\Sigma}_2^k+\hat{\Sigma}_2^k)\rVert_{\ast},\\	
&Var_{\beta} = \sigma^2_1 \left\lVert(n_1\hat{\Sigma}_1)^{-1}-(n_1\hat{\Sigma}_1+\hat{\Sigma}_2^k)^{-1}\right\rVert_{\ast}.
\end{align}\end{small}
\end{lemma}
\noindent {\it Remarks:} The above result bounds the increase in bias and the reduction in variance (see discussion of Figure \ref{Explanationtradeoff}).
Since our goal is to test MSE reduction --- in principle, 
we can use bootstrapping to calculate MSE approximately. 
This procedure has a significant computational footprint. 
Instead, \eqref{biasupper} (from a one-step Cauchy-Schwartz inequality), 
gives a {\em sufficient condition for MSE reduction} as shown below. 
\begin{theorem}\label{theorem:test}
{\bf a)}
Model \eqref{Ksetsmodel} has smaller MSE of $\hat{\beta}$ than model \eqref{Onesetmodel} whenever
\begin{equation}\label{eq:suffcond}
  H_0: \lVert G^{-1/2}\Delta\beta\rVert^2_2\leq \sigma^2_1.
\end{equation} 
    {\bf b)} Further, we have the following test statistic, 
\begin{equation}\label{eq:betatohat}
\left\lVert \frac{G^{-1/2}\Delta\hat{\beta}}{\sigma_1}\right\rVert^2_2\sim \chi^2_{(k-1)*p}\left(\left\lVert \frac{G^{-1/2}\Delta\beta}{\sigma_1}\right\rVert^2_2\right),
\end{equation}
where $\lVert G^{-1/2}\Delta\beta/\sigma_1\rVert_2$ is called a ``condition value''.
\end{theorem}
\noindent {\it Remarks:}
This is our main test result. Although $\sigma_i$ is typically unknown, it can be easily replaced using its site-specific estimation. 
Theorem \ref{theorem:test} implies that we can conduct a non-central $\chi^2$ distribution test based on the statistic.
Also, \eqref{eq:suffcond} shows that the non-central $\chi^2$ distribution, which the test statistics will follow, has a non-central parameter smaller than 
$1$ when the sufficient condition $H_0$ holds. 
Meanwhile, in obtaining the (surprisingly simple) sufficient condition $H_0$, no other arbitrary assumption is needed 
except the application of Cauchy-Schwartz inequality. 
From a practical perspective, Theorem \ref{theorem:test} implies that the sites, in fact, do {\em not} even need to share the full dataset 
to assess whether pooling will be useful.
Instead, the test only requires {\it very high-level} information such as $\hat{\beta}_i$, $\hat{\Sigma}_i$, $\sigma_i$ and $n_i$ for all participating 
sites -- 
which can be transferred very cheaply with no additional cost of data storage, or privacy implications. 
The following result deals with the special case where we have two participating sites. 
\begin{corollary}
  For the case where we have two participating sites, the condition \eqref{eq:suffcond} from Theorem \ref{theorem:test} reduces to
\begin{align} \label{eq:twositetest}
H_0:  \Delta \beta^T((n_1\hat{\Sigma}_1)^{-1}+(n_2\tau^2_2\hat{\Sigma}_2)^{-1})^{-1}\Delta \beta\leq \sigma^2_1.
\end{align}
\end{corollary}
\noindent {\it Remarks:}
The left side above relates to the Mahalanobis distance between $\beta_1$, 
$\beta_2$ with covariance $(n_1\hat{\Sigma}_1)^{-1}+(n_2\tau^2_2\hat{\Sigma}_2)^{-1}$, 
implying that the test statistic is a type of a normalized metric between the two regression models.

%%%%%%%%%%%%%%%%%%%%%%%%%%%%%%%%%%%%%%%%%%%%%%%%%%%%%%%%%%%%%%%%%%%%%%%%%%%%%%%%%%%%%%%%%%%%%%%%%%%%%%%%%%%%%%%%%%%%%%%%

\subsection{Sharing a subset of \mathversion{bold}$\beta$s\mathversion{bold}} \label{sec:subset}

\begin{figure}[!b]
 \vskip -0.2in
  \begin{center}
    %\fbox{
      \begin{tikzpicture}
      \draw 
      (-3,0) node[circle, draw, very thick, red, scale = 0.8](y1) {$Y_1$} 
      (-0.5,0.5) node[circle, draw, very thick, red, scale = 0.8](x1) {$X_1$}
      (-1.2, -0.5) node[circle, draw, very thick, red, scale = 0.8](z1){$Z_1$}
      (3,0) node[circle, draw, very thick, blue, scale = 0.8](y2) {$Y_2$} 
      (0.5,0.5) node[circle, draw, very thick, blue, scale = 0.8](x2) {$X_2$}
      (1.2, -0.5) node[circle, draw, very thick, blue, scale = 0.8](z2){$Z_2$};
      \draw[->, above, very thick](x1) -- node{$\beta$}(y1);
      \draw[->, below, very thick](z1) -- node{$\gamma_1$}(y1);
      \draw[->, above, very thick](x2) -- node{$\beta$}(y2);
      \draw[->, below, very thick](z2) -- node{$\gamma_2$}(y2);
      \draw
      (0.9,0.9) node {}
      -- (-0.9,0.9) node {}
      -- (-1.9,-0.9) node {}
      -- (1.9,-0.9) node {}
      -- cycle;
      %\draw[dotted, red, thick] \boundellipse{-1.5,0}{2}{1};
      %\draw[dotted, bluek, thick] \boundellipse{1.5,0}{2}{1};
      \draw [rounded corners = 40pt,red, dotted, thick, fill=red, opacity=0.2] (-4.3,0)--(0.4,1.3)--(-0.7,-1.4)--cycle;
      \draw [rounded corners = 40pt,blue, dotted, thick, fill=blue, opacity=0.2] (4.3,0)--(-0.4,1.3)--(0.7,-1.4)--cycle;
    \end{tikzpicture}
    %}
  \end{center}
  \vspace*{-20pt}
  \caption{\footnotesize $X$ and $Z$ influence the response $Y$. After adjustment, $X_1$ and $X_2$ may be close requiring same $\beta$. 
    %The effects of them are expected same on $Y_1,Y_2$ and  we assume same $\beta$ for them.
    However, $Z_1$ and $Z_2$ may differ a lot, and we need different $\gamma_1$ and $\gamma_2$.}
  \label{ExplanationConfound}
  %\vskip -0.2in
\end{figure}
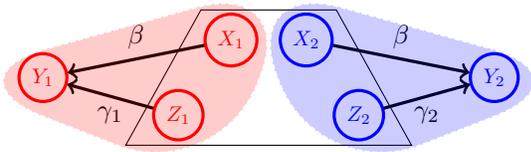

In numerous pooling scenarios, we are faced with certain systemic differences in the way predictors and responses associate across sites. 
For example, socio-economic status may (or may not) have a significant association
with a health outcome (response) depending on the country of the study (e.g., due to insurance coverage
policies). 
Unlike in Section \ref{sec:all}, we now relax the restriction that all coefficients are the same across sites, see Fig. \ref{ExplanationConfound}.
The model in \eqref{Ksetsmodel} will now include another design matrix of predictors $Z\in R^{n\times q}$ and corresponding coefficients $\gamma_i$ for each site $i$,
\begin{align}
  \label{Confoundingmodel}
  \min_{\beta,\gamma}\quad & \sum_{i=1}^k \tau^2_i\lVert y_i-X_i\beta-Z_i\gamma_i\rVert^2_2\\
  y_i&=X_i\beta^{\ast} + X_i\Delta \beta_i +Z_i\gamma^{\ast}_i+\epsilon_i,\ \tau_1=1
\end{align}

Our goal is still to evaluate whether the MSE of $\beta$ reduces.
We do not take into account the MSE change in $\gamma$ because they correspond to site-specific variables.
For estimation, $\hat{\beta}$ can first be computed from \eqref{Confoundingmodel}.
Treating it as a fixed entity now, $\hat{\gamma}_i$ can be computed using $y_i$ and $Z_i$ on each site independently.
Clearly, if $\hat{\beta}$ is close to the ``true'' $\beta^{\ast}$,
it will also enable a better estimation of site-specific variables.
It turns out that, if $\hat{\Sigma}_i$s are replaced by the conditional covariance, 
the analysis from Section \ref{sec:all} still holds for this case. 
Specifically, let $\hat{\Sigma}_{ab_i}$ be the sample covariance matrix
between features $a$ and $b$ from some site $i$. We have, 
\vspace{-2pt}
\begin{theorem}\label{theorem:subset}
	Analysis in Section \ref{sec:all} holds for $\beta$ in \eqref{Confoundingmodel} by replacing $\hat{\Sigma}_i$ with $\tilde{\Sigma}_i=\hat{\Sigma}_{xx_i}-\hat{\Sigma}_{xz_i}(\hat{\Sigma}_{zz_i})^{-1}\hat{\Sigma}_{zx_i}$
\end{theorem}
\vspace{-2pt}
\noindent {\it Remarks:}
The test now allows evaluating power improvements focused only on the subset of coefficients that is shared
and permits site-specific confounds. 
For example, we can test which subset of parameters might benefit from parameter estimation on pooled data from multiple sites.

\section{Pooling in High Dimensional Regression}\label{sec:lasso}
We now describe our analysis of pooling multi-site data in the high-dimensional setting where $p \gg n$.
The challenge here is that {\em variable section} has to be a first order concern. 
In classical regression, $\ell_2$ consistency properties are well known 
and so our focus in Section \ref{sec:hyptest} was devoted to deriving sufficient conditions for the hypothesis test. 
In other words, imposing the same $\beta$ across sites works in \eqref{Ksetsmodel} because we understand its consistency. 
In contrast, here, one cannot enforce a shared $\beta$ for all sites {\it before} the active set of predictors within each site are selected ---
directly imposing the same $\beta$ leads to a loss of $\ell_2$-consistency, making follow-up analysis problematic. 
Therefore, once a suitable model for high-dimensional multi-site regression is chosen, the first requirement is to characterize
its consistency. 

We start with the multi-task Lasso (a special case of group Lasso) \cite{liuml09}, where the authors show that the strategy selects better explanatory features compared to separately fitting Lasso on each site. 
But this algorithm underperforms when the sparsity pattern of the predictors is not identical across sites, so 
we use a recent variant called sparse multi-task Lasso \cite{lee10} -- essentially substituting ``sites'' for ``tasks''. 
The sparse {\it multi-site} Lasso in $p \gg n$ setting ($p$ is the number of predictors) is given as 
\begin{align} \label{eq:sp-mul-tak-lasso}
  \hat{B}^{\lambda} &=& \arg\min_{\beta} \sum_{i=1}^k \lVert y_i-X_i\beta_i\rVert^2_2 + \lambda \Lambda(B)\\
  \Lambda(B)&=&\alpha\sum_{j=1}^p\lVert\beta^j\rVert_1+(1-\alpha)\sqrt{k}\sum_{j=1}^p\lVert\beta^j\rVert_2, 
\end{align}
where $\lambda$ is the Lasso regularization parameter.
Here, $B\in R^{k\times p}$ is a matrix where the $i^{th}$ row gives the coefficients from $i^{th}$ site ($k$ sites in total). 
Also, $\beta_i$ with subscript denotes the $i^{th}$ row (site) of $B$, we use $\beta^j$ with superscript to give
the $j$-th column (coefficients) of $B$. 
The hyper-parameter $\alpha \in [0,1]$ balances the two penalties (and will be used shortly); 
a larger $\alpha$ weighs the $\ell_1$ penalty more and a smaller $\alpha$ puts more weight on the grouping.
%This will play an important role for the remainder of this section.
Similar to a Lasso-based regularization parameter, $\lambda$ here will produce a solution path (to select coefficients) for a given $\alpha$.
We first address the consistency behavior of the sparse multi-site Lasso in \eqref{eq:sp-mul-tak-lasso}, which was not known in the literature. 

\subsection{$\ell_2$ consistency} \label{sec:consis}

Our analysis of \eqref{eq:sp-mul-tak-lasso} is related to known results for Lasso \cite{mein09} and the group Lasso \cite{liu09}. 
Recall that $X_1,\ldots,X_k$ are the data matrices from $k$ sites.
We define $\bar{n}=\max_{i=1}^k\{n_i\}$ and $C = \bar{n}^{-1} \textsc{diag}(X^T_1X_1,...,X^T_kX_k)$
where $\textsc{diag}(A, B)$ corresponds to constructing a block-diagonal matrix with $A$ and $B$ as blocks on the diagonal.
We require the following useful properties of $C$ ($\lVert \cdot \rVert_0$ denotes $\ell_0$-norm).
\begin{definition}\label{def:msparse}
  The $m$-sparse minimal and maximal eigenvalues of $C$, denoted by $\phi_{\min}(m)$ and $\phi_{\max}(m)$, are
  {\begin{equation}
\min_{\nu:\lVert \nu\rVert_{0}\leq \lceil m\rceil}\frac{\nu^TC\nu}{\nu^T\nu}\quad\text{and}\quad
\max_{\nu:\lVert \nu\rVert_{0}\leq \lceil m\rceil}\frac{\nu^TC\nu}{\nu^T\nu}
\end{equation}}
\end{definition}

We call a feature ``active'' if its coefficient is non-zero.
Now, each site may have different active features: let $s_h \leq kp$ be the sum of the number of active features
  over all sites. Similarly, 
  $s_p$ is the cardinality of the union of features that are active in at least one site ($s_h \leq k s_p, s_p \leq p$). 
  Recall that when $\alpha \neq 0$, we add the Lasso penalty to the multi-site Lasso penalty.
When the sparsity patterns are assumed to be similar across all sites, $\alpha$ is small.
In contrast, to encourage site-specific sparsity patterns, we may set $\alpha$ to be large.
We now analyze these cases independently.
\begin{theorem} \label{theorem:grouplassotype}
Let $0\leq \alpha\leq 0.4$. Assume there exist constants $0\leq \rho_{\min}\leq \rho_{\max}\leq \infty$ such that
\begin{small} \begin{equation}\begin{aligned}
&\liminf_{n\rightarrow \infty}\phi_{\min}\left(s_p\left(1+\frac{2\alpha}{1-2\alpha}\right)^2\right)\geq \rho_{\min}\\
&\limsup_{n\rightarrow \infty}\phi_{\max}(s_p+\min\{\sum_{i=1}^k n_i,kp\})\leq \rho_{\max}.
\end{aligned}\end{equation} \end{small}
Then, for $\lambda\propto\sigma\sqrt{\bar{n}\log(kp)}$, there exists a constant $\omega>0$ such that, with probability converging to 1 for $n\rightarrow \infty$,
\begin{equation}\label{eq:refback}
\frac{1}{k}\lVert\hat{B}^{\lambda}-B^\ast\rVert^2_F\leq \omega\sigma^2\frac{\bar{s}\log (kp)}{\bar{n}},
\end{equation}
where $\bar{s}=\{(1-\alpha)\sqrt{s_p}+\alpha\sqrt{s_h/k}\}^2$, $\sigma$ is the noise level.
\end{theorem}
\noindent {\it Remarks:}
The above result agrees with known results for multi-task Lasso \cite{liuml09,liugl09} when the sparsity patterns are the same across sites. 
The simplest way to interpret Theorem \ref{theorem:grouplassotype} is via the ratio $r = \frac{s_h}{s_p}$.
Here, $r=k$ when the sparsity patterns are the same across sites. 
As $r$ decreases, the sparsity patterns across sites start to differ, in turn, 
the sparse multi-site Lasso from \eqref{eq:sp-mul-tak-lasso} will provide stronger consistency compared to the
multi-site Lasso (which corresponds to $\alpha = 0$).
In other words, whenever we expect {\emph site-specific} active features, 
the $\ell_2$ consistency of \eqref{eq:sp-mul-tak-lasso} will improve as one includes an additional
$\ell_1$-penalty together with multi-site Lasso.

Observe that for the non-sparse $\beta^j$, we can verify
that $\lVert\beta^j\rVert_1$ and $\sqrt{k}\lVert\beta^j\rVert_2$ have the same scale. 
On the other hand, for sparse $\beta^j$, $\lVert\beta^j\rVert_1$  has the same scale as $\lVert\beta^j\rVert_2$, i.e., with no $\sqrt{k}$ penalization
(see appendix). 
Unlike Theorem \ref{theorem:grouplassotype} where the sparsity patterns across sites are similar, 
due to this scaling issue, the parameters $\alpha$ and $\lambda$ need to be `corrected' for the setting where sparsity patterns have little overlap. 
We denote these corrected versions by $\tilde{\alpha} = \frac{\alpha}{(1-\alpha)\sqrt{k}+\alpha}$ and $\tilde{\lambda}=((1-\alpha)\sqrt{k}+\alpha)\lambda$. 
\begin{theorem}	\label{theorem:Lassotype}
Let $0.4\leq \tilde{\alpha}\leq 1$. Assume there exist constants $0\leq \rho_{\min}\leq \rho_{\max}\leq \infty$ such that
\begin{small} \begin{equation}\begin{aligned}
&\liminf_{n\rightarrow \infty}\phi_{\min}\left(s_h\left(1+\frac{(1-\tilde{\alpha})}{\tilde{\alpha}}\right)^2\right)\geq \rho_{\min}\\
&\limsup_{n\rightarrow \infty}\phi_{\max}(s_h+\min\{\sum_{i=1}^k n_i,kp\})\leq \rho_{\max}.
\end{aligned}\end{equation} \end{small}
Then, for $\tilde{\lambda}\propto\sigma\sqrt{\bar{n}\log(kp)}$, there exists  $\omega>0$ such that, with probability converging to $1$ for $n\rightarrow \infty$,
we have \eqref{eq:refback} with 
$\tilde{s}=\{ (1-\tilde{\alpha})\sqrt{s_p/k}+\tilde{\alpha}\sqrt{s_h/k} \}^2$ instead of $\bar{s}$.
\end{theorem}
\noindent {\it Remarks:} 
This result agrees with known results for Lasso \cite{mein09} when the sparsity patterns are completely different across sites. 
In this case (i.e., $\alpha$ is large),
the sparse multi-site Lasso has stronger consistency compared to Lasso ($\alpha = 1$).
The sparse multi-site Lasso is preferable as $r= \frac{s_h}{s_p}$ increases.
Note that although $\tilde{\alpha}$ and $\tilde{\lambda}$ are used for the results instead of $\alpha$ and $\lambda$, 
in practice, one can simply scale the chosen $\alpha$s appropriately, 
e.g., with $k=100$, we see that $\alpha \approx 0.99$ corresponds to $\tilde{\alpha}=0.95$.

\paragraph{\bf Performing hypothesis tests:} \label{high-dim-hyp}
Theorems \ref{theorem:grouplassotype} and \ref{theorem:Lassotype} show consistency of sparse multi-site Lasso estimation.
Hence, if the hyper-parameters $\alpha$ and $\lambda$ are known, we can estimate the coefficients $B^\ast$. 
This variable selection phase can be followed by a hypothesis test, similar to 
Theorem \ref{theorem:test} from Section \ref{sec:hyptest}.
The only remaining issue is the choice of $\alpha$. The existing methods
show that joint cross-validation for $\alpha$ and $\lambda$ performs unsatisfactorily and instead use a heuristic: 
set it to $0.05$ when it is known that sparsity patterns are similar across sites and $0.95$ otherwise \cite{simon13}.
%Joint cross-validation for $\alpha$ and $\lambda$ is shown to perform worse \cite{simon13}. {\color{red}They show joint cross-validation for $\alpha$ and $\lambda$ perform worse and set $\alpha$ to $0.05$ when it is known that sparsity patterns are similar across sites and $0.95$ otherwise \cite{simon13}}
Below, instead of a fixed $\alpha$, we provide a data-driven alternative that works well in practice. 

\paragraph{\bf Choosing \mathversion{bold}$\alpha$\mathversion{normal} using simultaneous inference:} \label{sec:choose}
Our results in Thm. \ref{theorem:grouplassotype} (and Thm. \ref{theorem:Lassotype} resp.) seem to suggest that increasing (and decreasing resp.) 
$\alpha$ will always improve consistency; however, this ends up requiring stronger $m$-sparsity conditions. We now describe a procedure 
to choose $\alpha$. 
First, recall that an active feature corresponds to a variable with non-zero coefficient.
We call a feature ``site-active'' if it is active at a site, 
an ``always-active'' feature is active at all $k$ sites. 
The proposed solution involves three steps. 
{\bf(1)} First, we apply simultaneous inference (like multi sample-splitting or de-biased Lasso) using all features 
at each of the $k$ sites with FWER control. 
This step yields ``site-active'' features for each site, and therefore, gives the set of always-active features and the sparsity patterns. 
{\bf(2)} Then, each site runs a Lasso and chooses a $\lambda_i$ based on cross-validation. 
We then set $\lambda_{\rm multi-site}$ to be the minimum among the best $\lambda$'s from each site. 
Using $\lambda_{\rm multi-site}$, we can vary $\alpha$ to fit various sparse multi-site Lasso models -- each run will select some number of always-active features.
We plot $\alpha$ versus the number of always-active features.
{\bf(3)} Finally, based on the sparsity patterns from the {\it site-active} set, we estimate whether the 
sparsity patterns across sites are similar or different (i.e., share few active features). Then, based on the plot from step (2), 
if the sparsity patterns from the site-active sets are different (similar) across sites, then the smallest (largest) value of $\alpha$ 
that selects the minimum (maximum) number of always-active features is chosen. The appendix includes details.

\section{Experiments} \label{sec:exps}

\begin{figure*}[ht]\centering %\vskip 0.2in
		\subfigure[][]{			%\label{Simulation-a}
			\includegraphics[width=0.63\columnwidth]{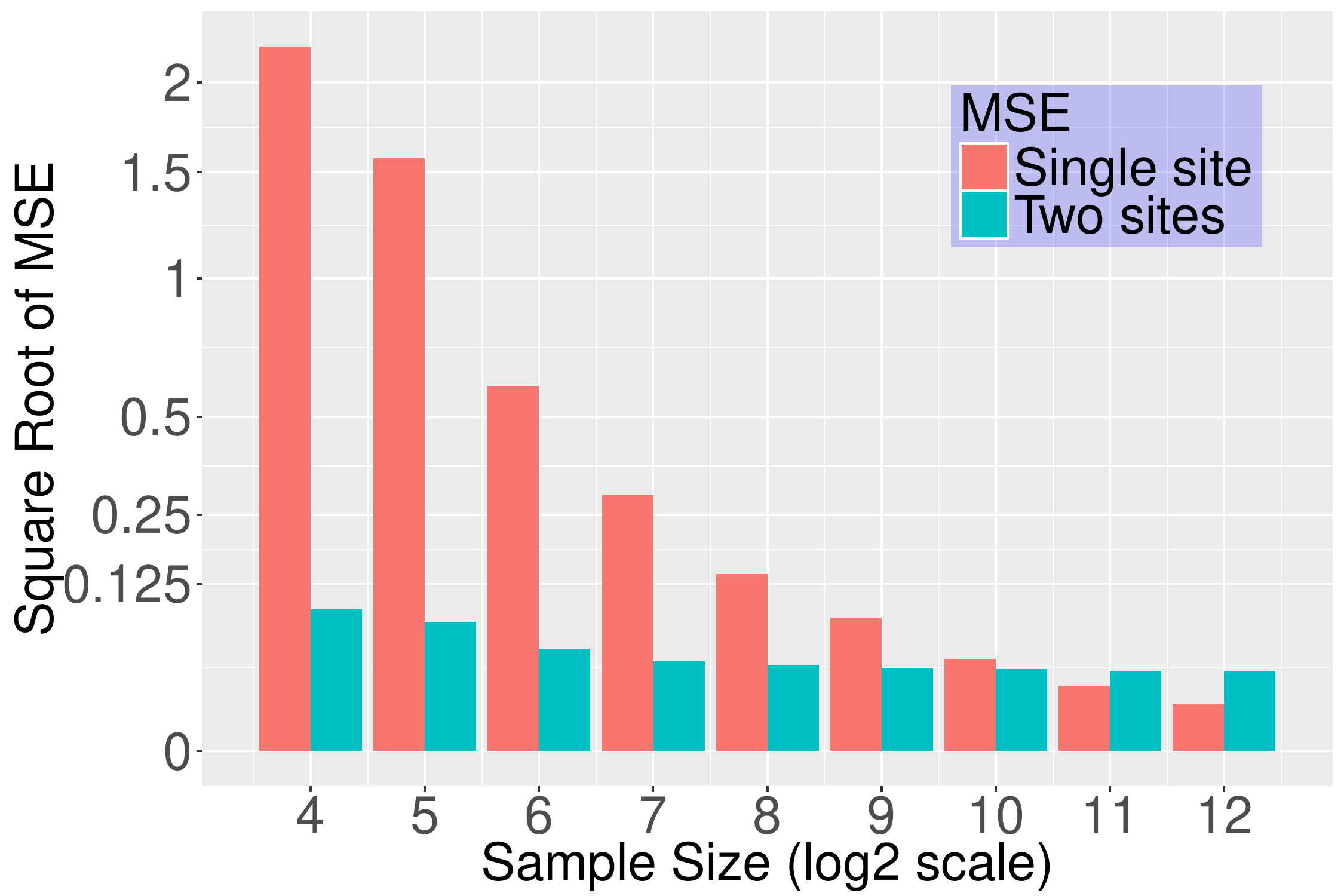}}\quad
		\subfigure[][]{			%\label{Simulation-b}
			\includegraphics[width=0.63\columnwidth]{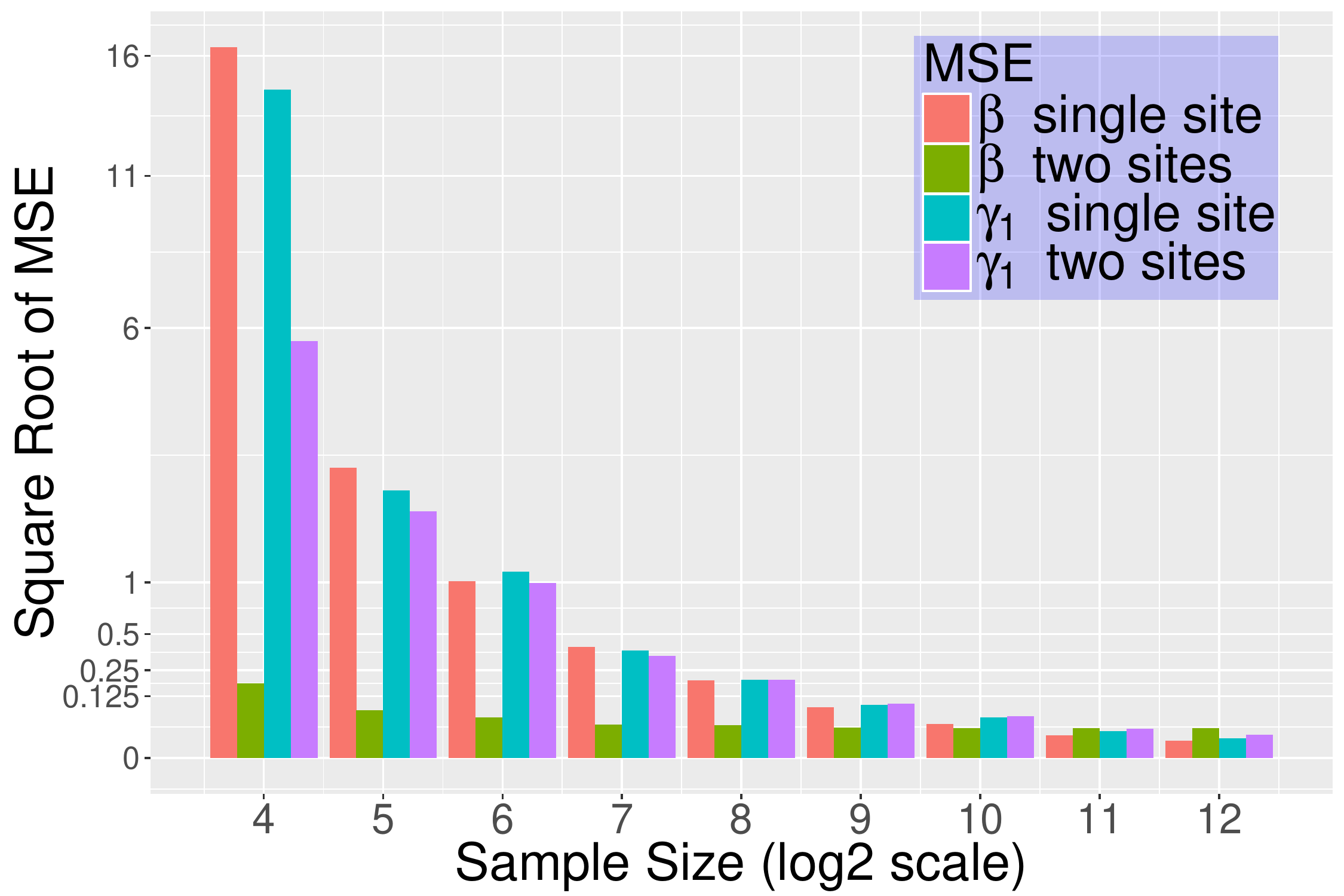}}\quad
		\subfigure[][]{			%\label{Simulation-c}
			\includegraphics[width=0.63\columnwidth]{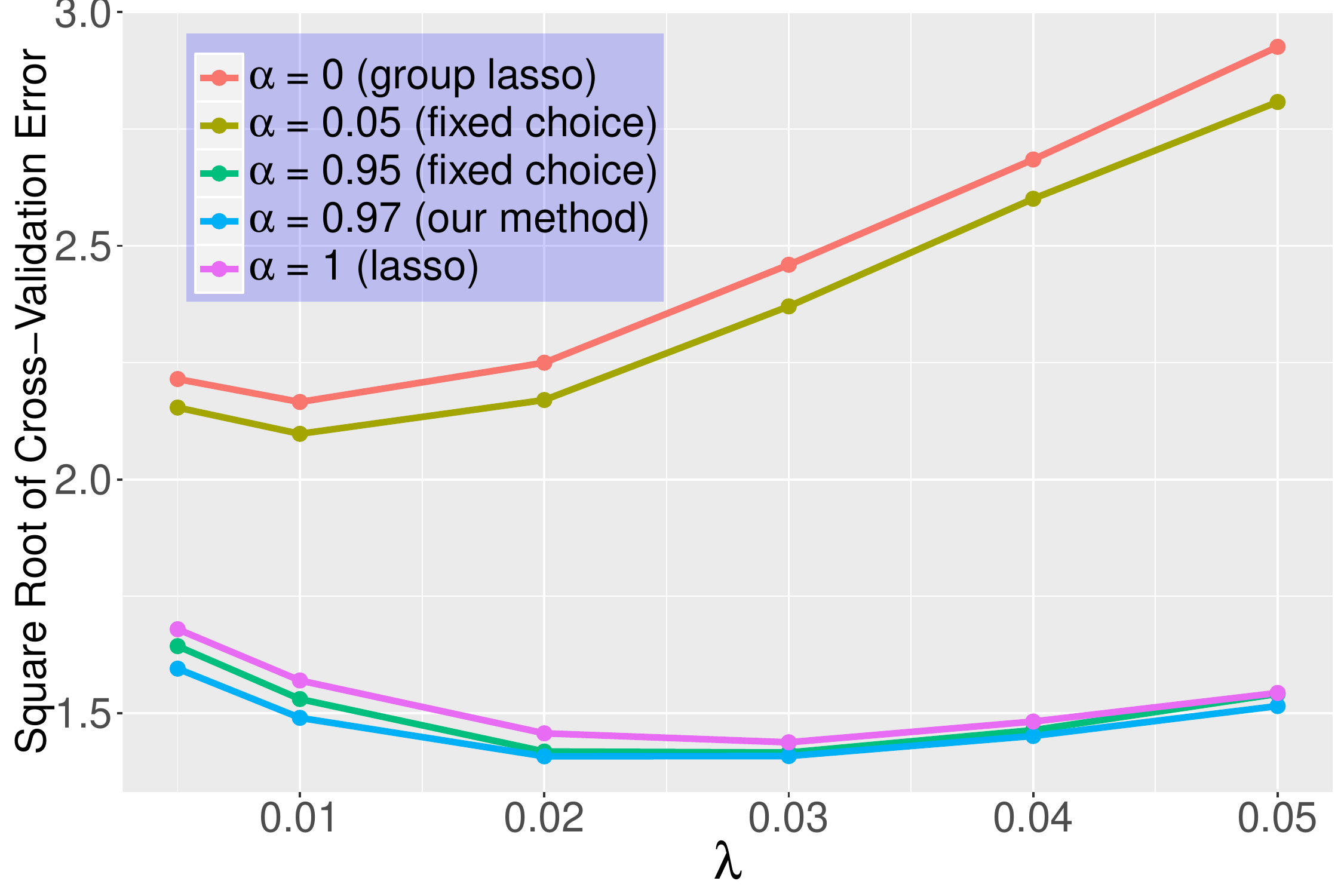}}\\ \vspace{-3mm}
		\subfigure[][]{			%\label{Simulation-d}
			\includegraphics[width=0.63\columnwidth]{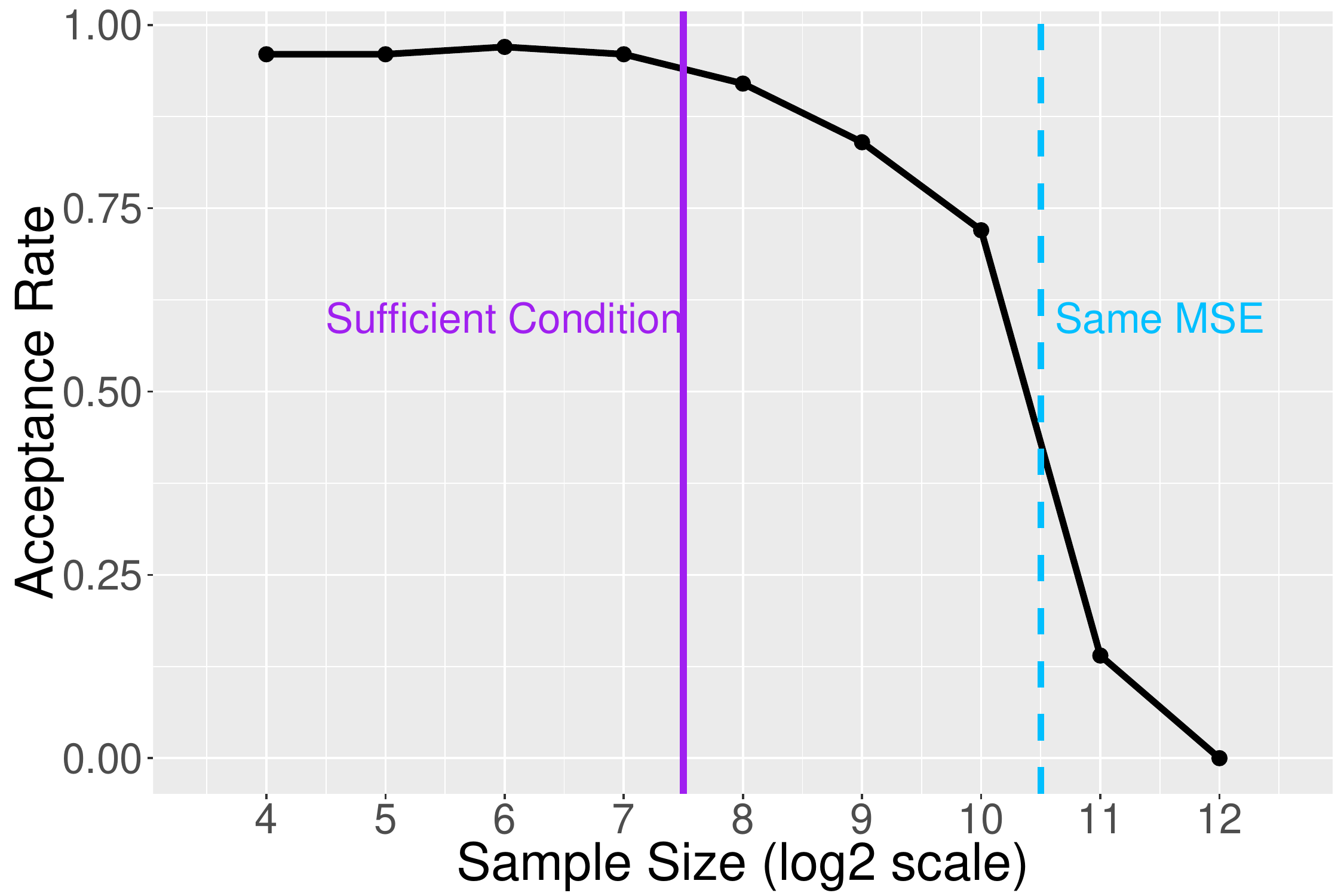}}\quad
		\subfigure[][]{			%\label{Simulation-e}
			\includegraphics[width=0.63\columnwidth]{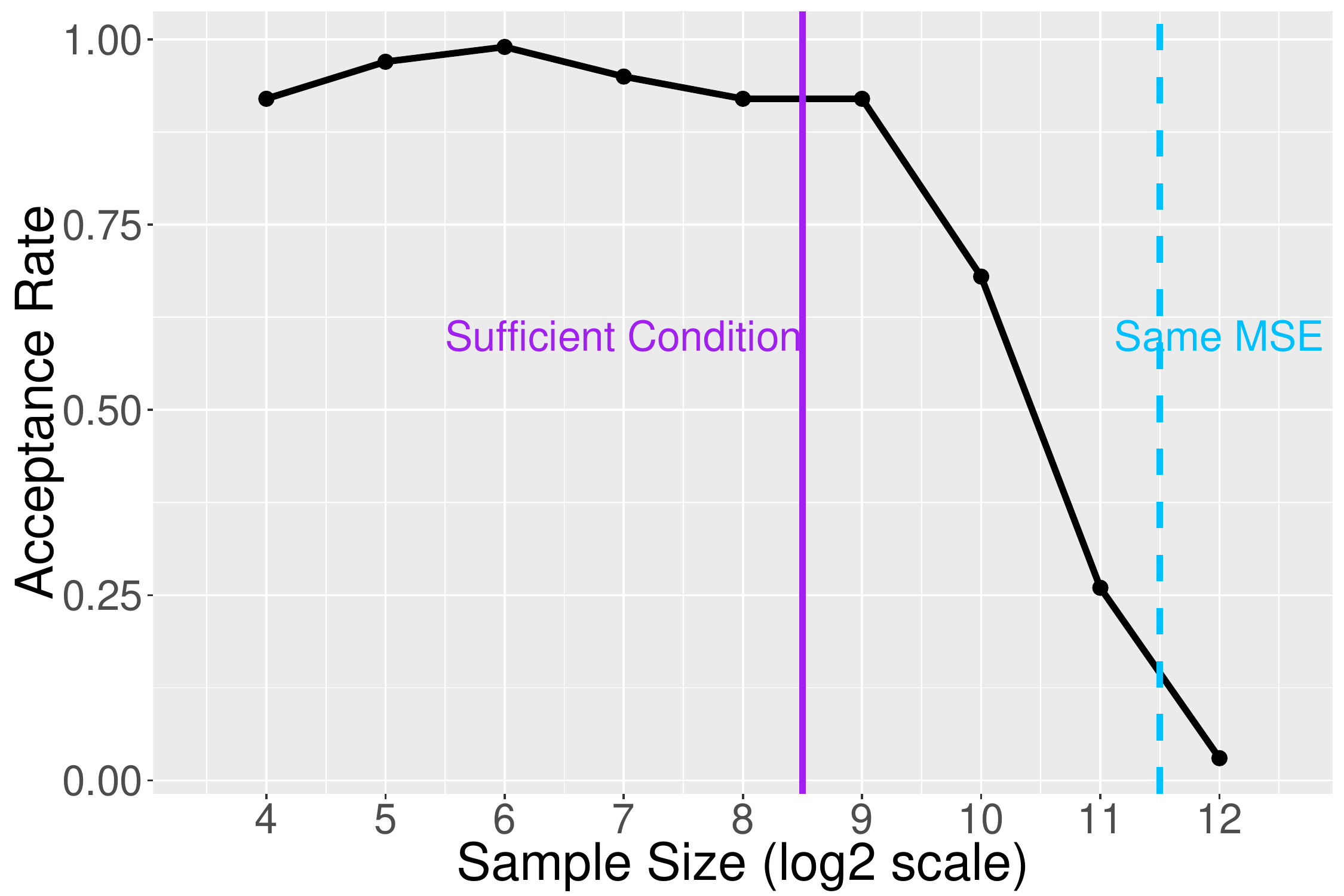}}\quad
		\subfigure[][]{			%\label{Simulation-f}
			\includegraphics[width=0.63\columnwidth]{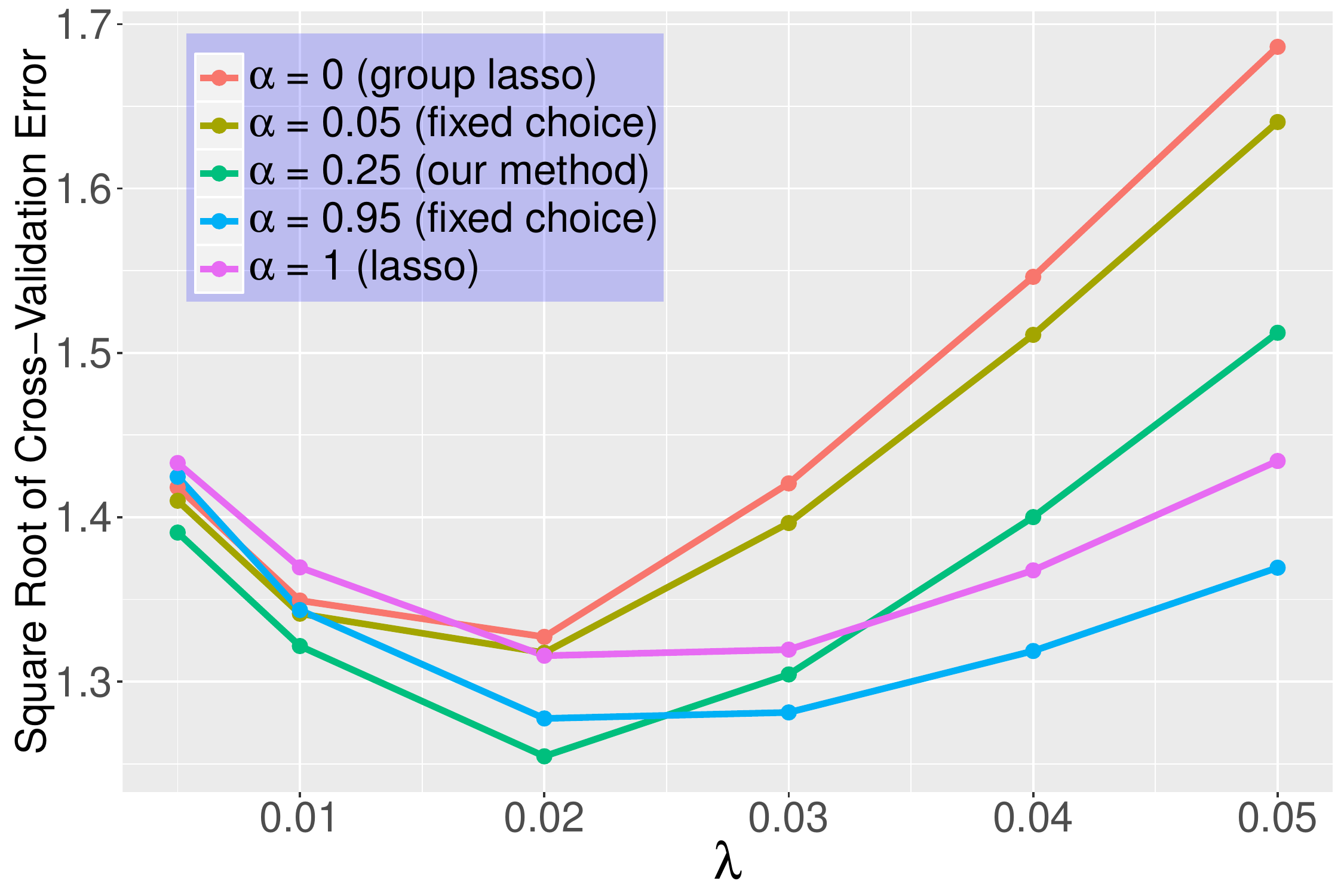}} \vspace{-13pt}
\caption{ \footnotesize \label{Simulation}  (a,d) $\hat{\beta}$'s MSE and the acceptance rate (Sec \ref{sec:all}), 
  (b,e) MSE of $\hat{\beta}$ and $\hat{\gamma}_1$, and the acceptance rate (Sec \ref{sec:subset}) using
  $100$ bootstrap repetitions.
  Solid line in (d,e) is when the condition from Theorem \ref{theorem:test} is $1$.
  Dotted line is when MSE of single-site and multi-site models are the same.
  (c) $\lambda$ error path when sparsity patterns are dissimilar across sites, %and group Lasso penalty is added to balance Lasso.
  (f) The regime where sparsity patters are similar.}%. (c,f) involve $10$-fold cross validation.}
\end{figure*}

Our experiments are two-fold.
First we perform simulations evaluating the hypothesis test from \S\ref{sec:hyptest} and sparse multi-site Lasso from \S\ref{sec:lasso}. 
We then evaluate pooling two Alzheimer's disease (AD) datasets from different studies to evaluate improvements in power,
and checking whether the proposed tests provide insights into the regimes when pooling is beneficial for regression,
and will yield tangible statistical benefits in investigating scientific hypotheses. 
\vspace{-3mm}
\paragraph{\bf Power and Type I Error of Theorem \ref{theorem:test}:} \label{sec:simall}
The first set of simulations evaluate the setting from Section \ref{sec:all} where the coefficients are same across two different sites.
The inputs for the two sites are set as $X_1$, $X_2 (\in \mathbb{R}^{n\times 3})\sim \mathcal{N}(0,\Sigma)$ with $\Sigma = 0.5(I+E)$ (where $I$ is identity and $E$ is a $3\times 3$ matrix of $1$s).
The true coefficients are given by $\beta_1 \sim U(0,4I)$ and $\beta_2 = \beta_1 + 0.1$ (where $U(\cdot)$ is multivariate uniform),
and the noise corresponds to $\epsilon_1\sim\mathcal{N}(0,3I)$ and $\epsilon_2\sim\mathcal{N}(0,0.5I)$ for the two sites respectively.
With this design, the responses are set as $y_1 = X_1\beta_1 + \epsilon_1$ and $y_2 = X_2\beta_2 + \epsilon_2$. 
Using $\{X_1,y_1\}$ and $\{X_2,y_2\}$, the {\it shared} $\hat{\beta}$ are estimated.
The simulation is repeated $100$ times with $9$ different sample sizes ($n=2^b$ with $b=4,\ldots,12$) for each repetition.
Fig.  \ref{Simulation}(a) shows the MSE of two-site (blue bars) and a baseline single-site (red bars) model computed using the corresponding $\hat{\beta}$s on site 1. 
Although both MSEs decrease as $n$ increases, the two-sites model consistently produces smaller MSE -- with large gains for small sample sizes (left-end of Fig.  \ref{Simulation}(a)).
Fig.  \ref{Simulation}(d) shows the acceptance rates of our proposed hypothesis test (from \eqref{eq:suffcond} and \eqref{eq:twositetest}) with $0.05$ significance level.
The purple solid line is the sufficient condition from Theorem \ref{theorem:test}, 
while the dotted line is where the MSE of the baseline single-site model starts to decrease below that of the two-site model.
The trend in Fig.  \ref{Simulation}(d) implies that as $n$ increases, the test tends to reject pooling the multi-site data with power $\rightarrow 1$.
Further, the type I error is well-controlled to the left of the solid line, and is low between the two lines.
See appendix for additional details about Figs. \ref{Simulation}(a,d).
\vspace{-3mm}
\paragraph{\bf Power and Type I Error of Theorem \ref{theorem:subset}:} \label{sec:simsubset}
The second set of simulations evaluate the confounding variables setup from Section \ref{sec:subset}.
Similar to Section \ref{sec:simall}, here we have $(X_1,Z_1)$, $(X_2,Z_2) \sim \mathcal{N}(0,\Sigma)$
with $\Sigma=\left(\begin{array}{cc}0.5 I_{3\times 3} +0.5 E_{3\times 3}, &0.2E_{3\times 5}\\0.2E_{5\times 3}, &0.8I_{5\times 5}+0.2E_{5\times 5} \end{array}\right)$.
$\beta_1$ and $\beta_2$ are the same as before. $\gamma_1=(1,1,2,2,2)^T$ and $\gamma_2=(2,2,2,1,1)^T$ are the coefficients for $Z_1$ and $Z_2$ respectively.
The new responses $y_1$ and $y_2$ will have the extra terms $Z_1\gamma_1$ and $Z_2\gamma_2$ respectively.
Fig.  \ref{Simulation}(b,e) shows the results. All the observations from Fig.  \ref{Simulation}(a,d) hold here as well.
For small $n$, MSE of two-site model is much smaller than baseline, and as sample size increases this difference reduces.
The test accepts with high probability for small $n$, and as sample size increases it rejects with high power. 
The regimes of low type I error and high power in Fig.  \ref{Simulation}(e) are similar to those from Fig.  \ref{Simulation}(d).

%%%%%%%%%%%%%%%%%%%%%%%%%%%%%%%%%%%%%%%%%%%%%%
\begin{figure*}[ht]\centering
	\subfigure[][]{
		\includegraphics[width=0.5\columnwidth]{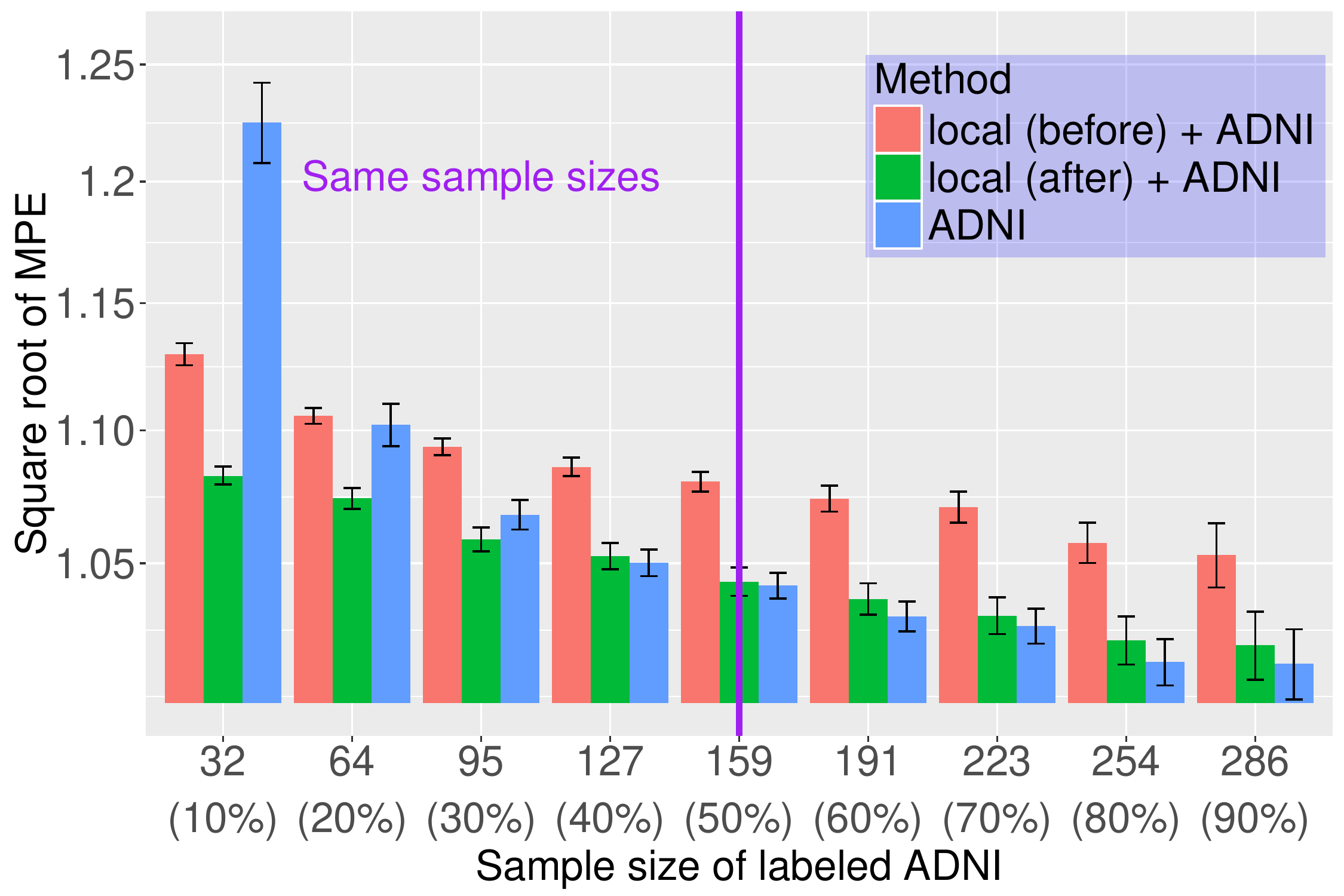}}
	\subfigure[][]{
		\includegraphics[width=0.5\columnwidth]{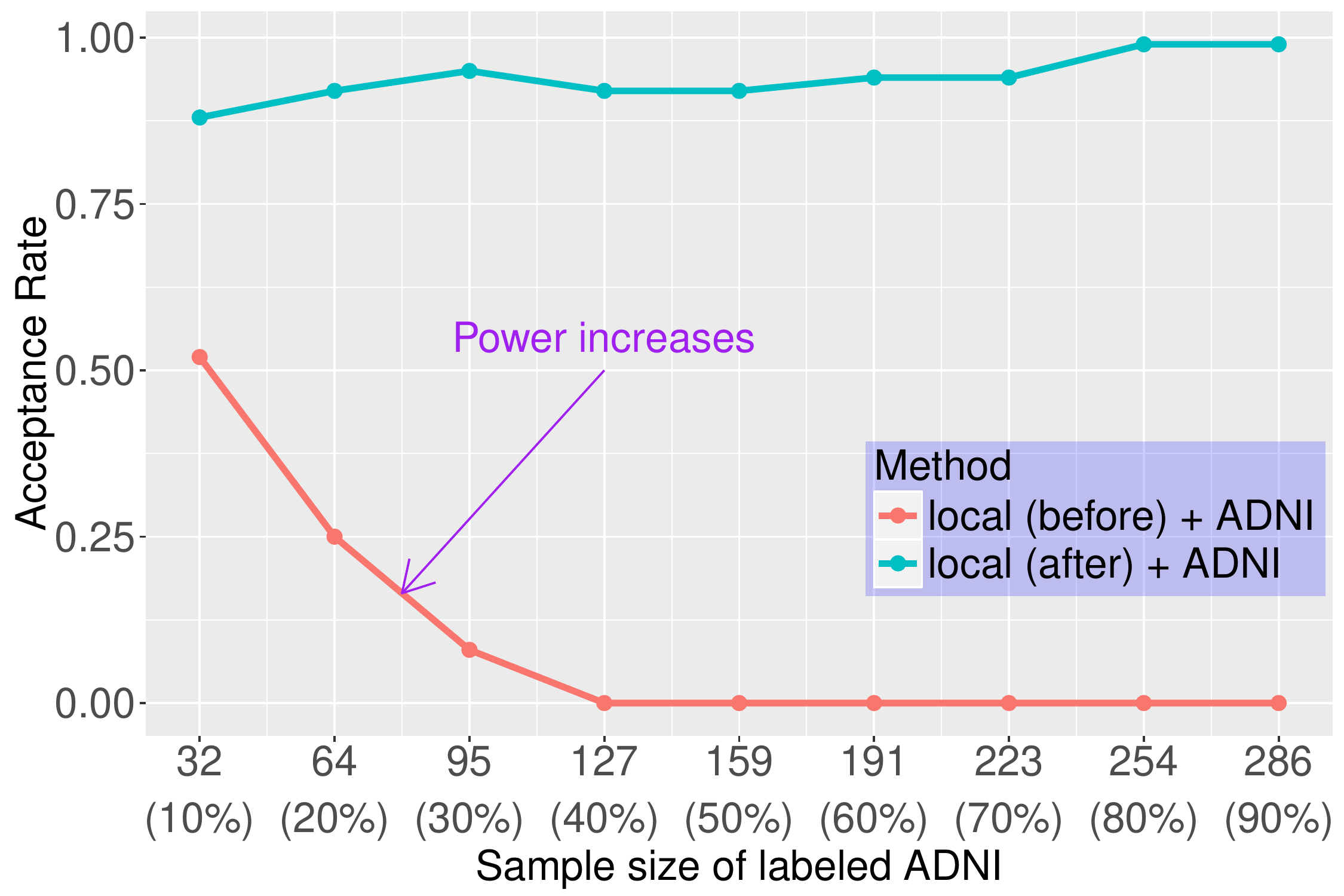}}
	\subfigure[][]{
		\includegraphics[width=0.5\columnwidth]{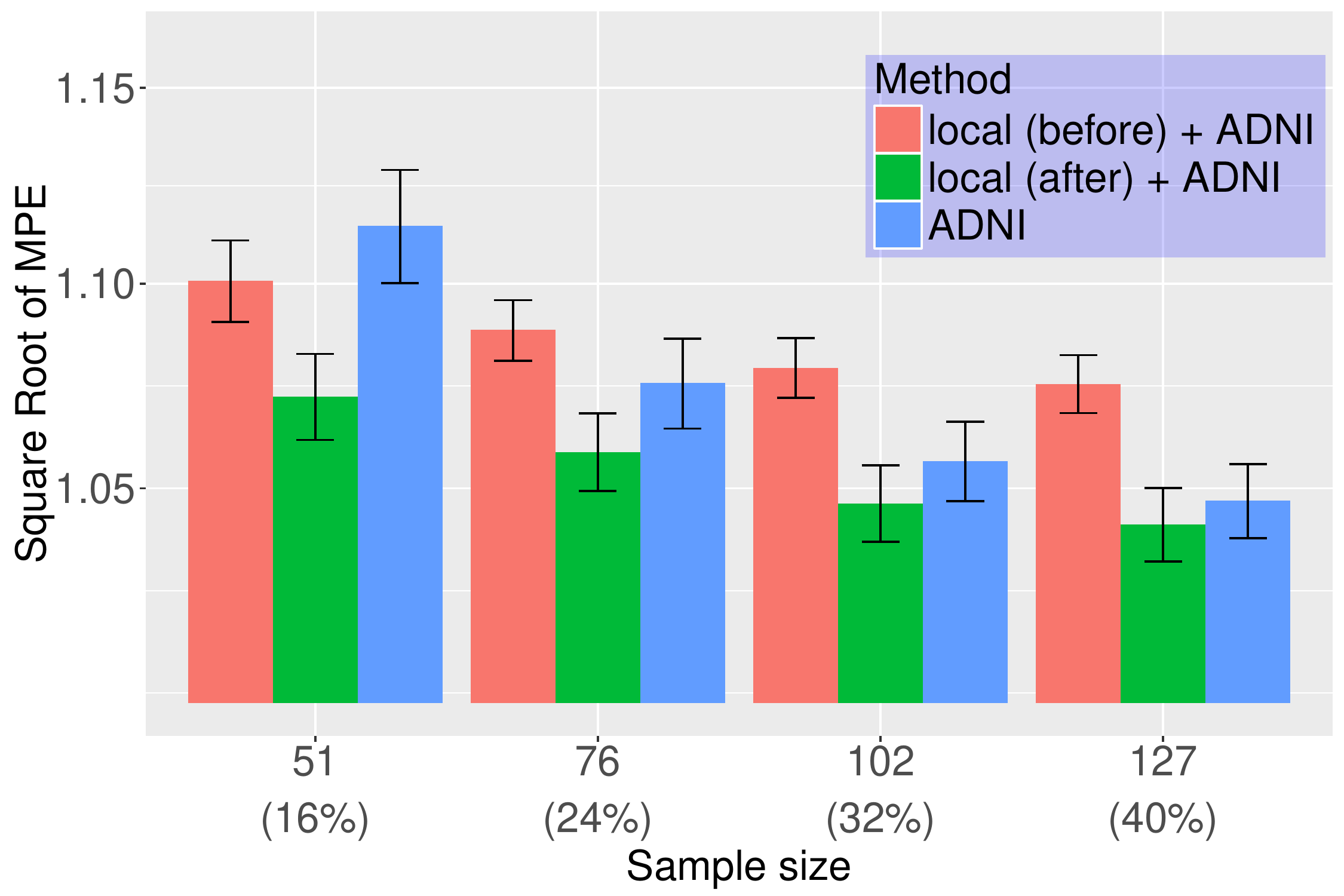}}
	\subfigure[][]{
		\includegraphics[width=0.5\columnwidth]{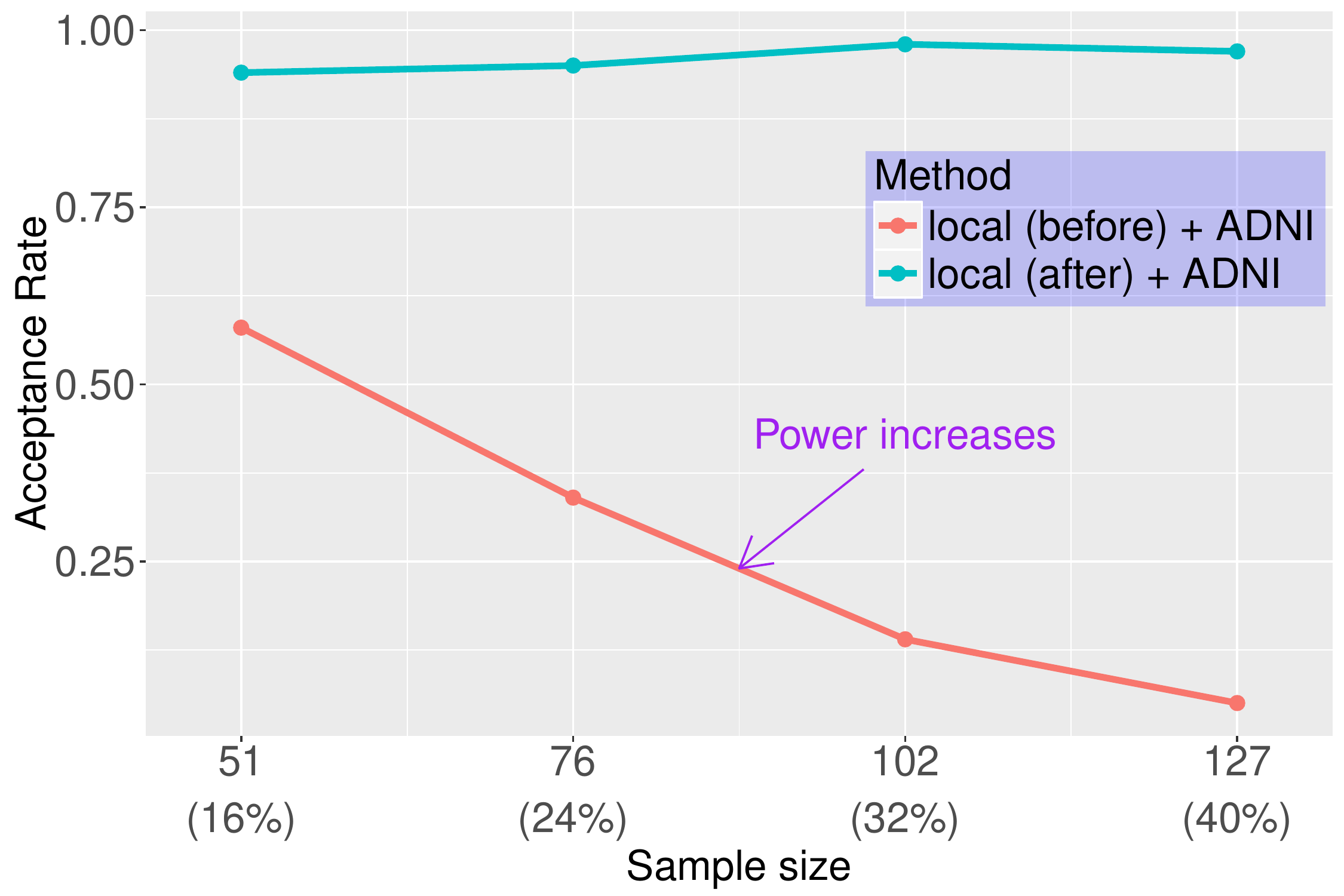}}\vspace{-13pt}
	\caption{\footnotesize  \label{Experiment}
          (a,c) MPE for the pooled regression model after/before  transformations (green/red) compared to baseline (blue) plotted against training subset size of ADNI.          
          $x$-axis is number/fraction of ADNI labeled samples used in training (apart from ADlocal).
          (b,d) show the acceptance rates for (a,c). %The purple line in (a) is where the training sample sizes of ADNI and ADlocal match.
          Unlike in (a), (c) restricts same training data size for ADNI and ADlocal.}
\end{figure*}

\subsection{Sparse multi-sites Lasso $\ell_2$-consistency}\label{sec:sec4exps} 

We now use $4$ sites with $n = 150$ samples each and $p = 400$ features to test the sparse multi-site model from \S \ref{sec:lasso}.
We set the design matrices $X_i$ ($i=1,\ldots,4$) $\sim \mathcal{N}(0,\Sigma)$ with $\Sigma = 0.8I_{p\times p}+0.2E_{p\times p}$. 
We consider the two cases (sparsity patterns shared/not shared) separately. 
\vspace{-3mm}
\paragraph{\bf Few sparsity patterns shared:} 
$6$ shared features and $14$ site-specific features (out of the $400$) are set to be active in $4$ sites. 
Each shared feature is sampled from $U(0,4)$ for the first two sites and $U(0,0.5)$ for the rest.
All the site-specific features are $\sim U(0,4)$.
The noise $\epsilon_i \sim \mathcal{N}(0,1)$, and the responses are $y_i=X_i\beta_i+\epsilon_i$.
Fig.  \ref{Simulation}(c) shows the $10$-fold cross validation error as $\lambda$ changes (i.e., solution path) for different $\alpha$ settings,
including the value from our proposed selection procedure (from Section \ref{sec:choose}), Lasso ($\alpha=1$),
group Lasso ($\alpha=0$) and arbitrary values $\alpha = 0.05$, $0.95$ (as suggested by \cite{simon13}).
Our chosen $\alpha=0.97$ (the blue curve in Fig.  \ref{Simulation}(c)) has the smallest error, across all $\lambda$s, thereby implying a better $\ell_2$ consistency.
Table $1$ in the appendix includes more details, including
$\alpha=0.97$ discovers more always-active features, while preserving the ratio of correctly discovered active features to all the discovered ones.
\vspace{-3mm}
\paragraph{\bf Most sparsity patterns shared:}
Unlike the earlier case, here we set $16$ shared and $4$ site-specific features (both $\sim U(0,4)$) to be active among all $400$ features. 
The result, shown in Fig.  \ref{Simulation}(f), is similar to Fig.  \ref{Simulation}(c).
The proposed choice of $\alpha = 0.25$ competes favorably with alternate choices while preserving the correctly discovered number of always-active features.
Unlike the previous case, the ratio of correctly discovered active features to all discovered ones increases here (see appendix).

%%%%%%%%%%%%%%%%%%%%%%%%%%%%%%%%%%%%%%%%%%%%%%

\subsection{Combining AD datasets from multiple sites}\label{sec:adcombine}

We now evaluate whether two AD datasets acquired at different sites --
an Alzheimer's Disease Neuroimage Initiative (ADNI) dataset and a local dataset from \href{http://www.adrc.wisc.edu/}{Wisconsin ADRC} (ADlocal) can be combined (appendix has dataset details).
The sample sizes are $318$ and $156$ respectively.
Cerebrospinal fluid (CSF) protein levels are the inputs, and the response is hippocampus volume.
Using $81$ age-matched samples from each dataset, we first perform domain adaptation (using a maximum mean discrepancy objective as a measure of distance between the two marginals),
and then transform CSF proteins from ADlocal to match with ADNI.
The transformed data is then used to evaluate whether adding ADlocal data to ADNI will improve the regression performed on the ADNI data.
This is done by training a regression model on the `transformed' ADlocal and a subset of ADNI data,
and then testing the resulting model on the remaining ADNI samples.
We use two baseline models each of which are trained using -- ADNI data {\em alone}; and {\it non-transformed} ADlocal (with ADNI subset).

Fig.  \ref{Experiment}(a,b) show the resulting mean prediction error (MPE) scaled by the estimated noise level in ADNI responses,
and the corresponding acceptance rate (with significance level $0.05$) respectively.
The $x$-axis in Fig.  \ref{Experiment}(a,b) represents the size of ADNI subset used for training. 
As expected, the MPE reduces as this subset size increases.
Most importantly, pooling after transformation (green bars) seems to be the most beneficial in terms of MPE reduction. 
As shown in Fig.  \ref{Experiment}(a), to the left of purple line where the subset size is smaller than ADlocal datasize, pooling the datasets improves estimation.
This is the small sample size regime which necessitates pooling efforts in general.
As the dataset size increases (to the right of $x$-axis in Fig.  \ref{Experiment}(a)) the resulting MPE for the pooled model is close to what we will achieve using the ADNI data by itself.

Since pooling after transformation is at least as good as using ADNI data alone, 
our hypothesis test accepts the combination with high rate ($\approx 95\%$), see Fig.  \ref{Experiment}(b).
The test rejects the pooling strategy with high power for combining before domain adaptation (see Fig.  \ref{Experiment}(b)), as one would expect.
This rejection power increases rapidly as sample size increases, see red curve in Fig.  \ref{Experiment}(b).
The results in Fig.  \ref{Experiment}(c,d) show the setting where one {\it cannot} change the dataset sizes at the sites i.e.,
the training set uses an equal number of labeled samples from both the ADNI and ADlocal ($x$-axis in Fig.  \ref{Experiment}(c)), 
and the testing set always corresponds to $20\%$ of ADNI data.
This is a more interesting scenario for a practitioner compared to Fig.  \ref{Experiment}(a,b),
because in Fig.  \ref{Experiment}(c,d) we use same sample sizes for both datasets.
The trends in Fig.  \ref{Experiment}(c,d) are the same as Fig.  \ref{Experiment}(a,b).

\vspace{-6pt}
\section{Conclusions} \label{sec:conc}

We present a hypothesis test to answer whether pooling multiple datasets acquired from different sites is guaranteed to increase statistical power for regression models. 
For both standard and high dimensional linear regression, we identify regimes where such pooling is sensible, and show how such policy decisions can be made via simple checks executable on each site before any data transfer ever happens. 
We also show empirical results by combining two Alzheimer's disease datasets in the context of different regimes proposed by our analysis, and see that the regression fit improves as suggested by the theory. The code is available at \href{https://github.com/hzhoustat/ICML2017}{https://github.com/hzhoustat/ICML2017}.

\vspace{-10pt}\noindent \paragraph{\bf Acknowledgments:} This work is supported by NIH grants R01 AG040396, R01 EB022883, UW CPCP AI117924, R01 AG021155, and NSF awards DMS 1308877, CAREER 1252725 and CCF 1320755.
The authors are grateful for partial support from UW ADRC AG033514, UW ICTR 1UL1RR025011 and funding from a UW-Madison/DZNE collaboration initiative.

%\newpage

\bibliography{Hao_ICML}
\bibliographystyle{icml2017}

\end{document}